\documentclass[prl,twocolumn,amsmath,amssymb,floatfix]{revtex4}
\usepackage{graphicx}
\usepackage{bm}

\newcommand{\be}{\begin{equation}}
\newcommand{\ee}{\end{equation}}
\newcommand{\beq}{\begin{eqnarray}}
\newcommand{\eeq}{\end{eqnarray}}

\begin{document}

\title{Shot noise of interference
between independent atomic systems.}

\author{Anatoli Polkovnikov}

\affiliation{Department of Physics, Boston University, Boston, MA
02215}

\begin{abstract}

We study shot (counting) noise of the amplitude of interference
between independent atomic systems. In particular, for the two
interfering systems the variance of the fringe amplitude decreases
as the inverse power of the number of particles per system with the
coefficient being a non-universal number. This number depends on the
details of the initial state of each system so that the shot noise
measurements can be used to distinguish between such states. We
explicitly evaluate this coefficient for the two cases of the
interference between bosons in number states and in broken symmetry
states. We generalize our analysis to the interference of multiple
independent atomic systems. We show that the variance of the
interference contrast vanishes as the inverse power of the number of
the interfering systems. This result, implying high signal to noise
ratio in the interference experiments, holds both for bosons and for
fermions.

\end{abstract}

\maketitle

Interference between independent systems is an interesting
phenomenon, which goes back to the discovery of the
Hanbury-Brown-Twiss (HBT) effect (see e.g.
Ref.~[\onlinecite{scully}]). The very possibility that particles
from independent condensates can interfere, i.e. have a certain
relative phase, is quite intriguing and is not entirely
obvious~\cite{anderson}. The origin of this phenomenon is the
quantum indistinguishability of identical particles and the
superposition principle. Recent experimental advances in cold atom
systems opened new possibilities to detect the interference both in
equilibrium~[\onlinecite{ketterle_int, zoran, stock, zoran_KT}] and
in nonequilibrium~\cite{schmiedmayer} situations. Such experiments
even allow probing various properties of interacting systems. For
example, an analysis of the scaling of the interference signal with
the imaging size lead to the direct observation of the
Kosterlitz-Thouless phase transition~\cite{zoran_KT}. The problem of
interference also attracted a lot of theoretical attention. In
Ref.~[\onlinecite{Javanainen}] Javanainen and Yoo numerically
studied the interference between two Fock states for a given set of
detectors. It was shown that the resulting count distribution of
atoms was similar to the one arising from the interference of two
condensates with randomly broken phases. Later Castin and Dalibard
analyzing another Gedanken experiment came to the same conclusion
and showed that these random phases are induced by the
detectors~\cite{dalibard}. Emerging interference between two number
states was later interpreted as an indication of the presence of
hidden variables~\cite{laloe}. In Refs.~[\onlinecite{pnas,gritsev}]
the authors studied the interference between independent fluctuating
condensates and reached a similar conclusion that for the large
ideal condensates the amplitude of interference does not fluctuate.
Recently Altman {\em et. al.} suggested using noise interferometry
as a new probe of interacting atomic systems~\cite{ehud_noise}.
These ideas were later experimentally
implemented~\cite{greiner_noise, bloch_noise}.

In fact the reason why two independent condensates with fixed number
of particles in each must interfere follows from the basic
principles of quantum mechanics. Indeed, because of the number-phase
uncertainty each of the condensates does not have a well defined
phase. However, according to quantum mechanics any measurement probe
sensitive to the phase difference (whether it is a time of flight
image or something else) will project the condensates to the state
with a well defined relative phase. As a result the condensates will
interfere but the relative phase will be random for each
experimental run in agreement with
Refs.~[\onlinecite{dalibard,pnas}] as well as this work.

The main purpose of this paper is to study shot noise of the
interference between independent atomic systems. For the most part
we will consider bosonic systems (condensates for short). However,
our approach is independent of the atom statistics and where
necessary we will give the explicit expressions for fermions as
well.

We point that if the systems are independent then the interference
between them can be considered as a noise, similarly to the HBT
effect. Indeed, the average of the atom density over many
experimental runs does not result in any oscillating component.
However, for the two condensates with large number of particles,
each run yields well defined interference fringes~\cite{Javanainen,
dalibard, laloe, pnas}. The fringe amplitude is thus a quantity
which does not average to zero. Moreover, as we mentioned above,
this amplitude is expected to have vanishing fluctuations.
Conversely, if the number of particles in each of the two systems is
small, we expect to see significant shot noise and large
fluctuations of the fringe amplitude. The analysis of these
fluctuations is the subject of the present investigation. On passing
on, we mention that shot noise experiments are a very powerful tool
in condensed matter physics (see Ref.~[\onlinecite{beenakker}] for a
review) and in quantum optics (see e.g. Ref.~[\onlinecite{scully}]).
Such experiments can be used to detect charge of quasi-particles,
transmission properties of small conductors, entanglement between
electrons, allow to distinguish quantum and classical light sources,
etc.

We emphasize that when one analyzes the interference between
extended systems, one encounters at least two different sources of
noise (apart from the noise associated with the probing beam). One
of them originates from the phase fluctuations within each system,
which can have either quantum or thermal origin. This type of noise
was analyzed in detail in Refs.~[\onlinecite{pnas, gritsev}]. In
particular, it was shown that the scaling of the fringe amplitude
with the system size and its distribution function contain
information about phase correlation functions within each system.
The second source of noise, which is studied in the present paper,
has a purely quantum nature originating from the commutation
relations between identical particles. This counting or shot noise
exists even in ideal noninteracting systems and as we show below it
contains important information about the nature of the interfering
states.

\begin{figure}[ht]
\includegraphics[width=8.5cm]{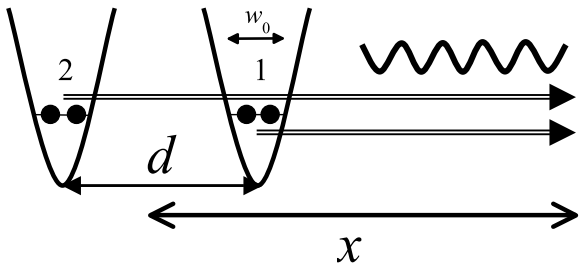}
\caption{Schematic view of the interference experiment between two
independent systems.}
\label{fig1}
\end{figure}

Assume that all bosons in each of the two interfering systems
systems ($1$ and $2$) occupy the same quantum state. We consider the
standard time of flight measurement setup (see Fig.~\ref{fig1}). At
a certain moment in time these particles are allowed to expand. For
simplicity we assume that the particles expand only in one direction
$x$. For the two dimensional systems with strong transverse
confinement like in e.g. Ref.~[\onlinecite{zoran_KT}] this
assumption is usually well justified since the transverse kinetic
energy is very large. For the system of two one-dimensional
condensates~\cite{schmiedmayer} the expansion for each condensate
occurs in the radial direction . In this case one can always analyze
the image in the plane of the two condensates and the present
analysis remains intact. In a more general case one has to compute
the overlap between the expanding Wannier functions corresponding to
the two different condensates. However, we emphasize that this
overlap brings a more complicated geometrical factor characterizing
the interference, which enters only as the prefactor into the
interference amplitude and does not affect main conclusions of this
paper. After waiting until the size of the clouds greatly exceeds
the original separation between the condensates $d$, the density of
these atoms is measured using probing beams. We assume that there is
no photon shot noise and the atoms are detected with a $100\%$
probability. For simplicity we also assume that $d$ is much larger
than the initial size of each condensate $w_0$.

As we already argued, there is no average interference contrast
between the two condensates. However, there is a well defined
observable $A_2$, equal to the square of the interference
amplitude~\cite{pnas}. Let us define the fluctuating variable $A$,
average of which over many experimental runs gives $A_2$, in the
following way:
\be
A=\int\int dx dx' p(x,t)p(x',t) e^{iQ(x-x')}-\int dx\, p(x,t),
\label{aq_2}
\ee
Here $p(x,t)$ is the number of the absorbed photons at the time $t$
at the position $x$ and $Q=md/\hbar t$, where $m$ is the atom's
mass. The integral over $x$ and $x'$ should be understood as the
sum, where the discretization step is determined by the detector and
can not be smaller than the photon wavelength. Apart from the second
term in Eq.~(\ref{aq_2}), which as we will see below has a quantum
origin, the expression for $A$ is just a square of the Fourier
transform of the absorbtion image taken at the wave vector $Q$. Note
that since by assumption all the detectors have $100\%$ efficiency,
the number of the absorbed photons coincides with the number of
atoms at the detection point: $p(x,t)=n(x,t)\equiv
a^\dagger(x,t)a(x,t)$, where $a^\dagger(x,t)$ and $a(x,t)$ are the
time-dependent creation and annihilation operators of the atoms. We
understand the equality sign here and in expressions below in the
sense that the statistical properties of $p(x,t)$ and the
corresponding quantum operator are equivalent.

Using the bosonic commutation relations
$[a(x,t),a^\dagger(x^\prime,t)]=\delta(x-x^\prime)$ it is easy to
see that Eq.~(\ref{aq_2}) can be rewritten as
\be
A=\int\int dx dx' a^\dagger(x,t)a^\dagger(x',t) a(x,t)
a(x',t)\mathrm e^{iQ(x-x')}.
\label{aq_4}
\ee
We can further simplify Eq.~(\ref{aq_4}) using that in the long time
limit~\cite{pit, zoran}
\be
a(x,t)\approx  a_1\,u(x,t)\mathrm e^{iQ_1(t) x}+ a_2\,u(x,t)\mathrm
e^{iQ_2(t) x},
\label{a(t)}
\ee
where $Q_{1,2}=m(x\pm d/2)/\hbar t$, $a_{1,2}$ are the bosonic
operators in the Schr\"odinger representation corresponding to the
systems $1$ and $2$, and $u(x,t)$ is the time-dependent Wannier
function corresponding to either of the two condensates (for
simplicity we assume that these Wannier functions are identical).
Using Eqs.~(\ref{aq_4}) and (\ref{a(t)}) we find
\be
A_2\equiv{\overline A}=\langle a_1^\dagger a_2^\dagger a_2
a_1\rangle,
\label{a2}
\ee
where the overline implies averaging over many experimental runs and
the angular brackets denote the expectation value. We note that if
the initial width of the condensates $w_0$ is finite then there will
be corrections to Eq.~(\ref{a2}) proportional to $\exp(-d^2/w_0^2)$.
However, they are negligible as long as $d\gg w_0$.

We would like to emphasize that it is crucial to first take the
normal order of the operator $A$ (or any other operator), like we
did in Eq.~(\ref{aq_4}), and only then use Eq.~(\ref{a(t)}). Indeed,
it is easy to check that the asymptotic form (\ref{a(t)}) does not
preserve the correct commutation relations between the operators
$a(x,t)$ and $a^\dagger(x^\prime,t)$. It is therefore very important
to order the operators first and use Eqs.~(\ref{a2}) only after
that. It is straightforward to check that if one proceeds in a
different way then one will get un-physical results with negative
probabilities for the observables.

One can similarly analyze the fermionic case. The resulting
expression for the interference amplitude is identical to
Eq.~(\ref{a2}). There is, however, one subtlety. Namely the mean
value of $A$ defined in Eq.~(\ref{aq_2}) is negative, reflecting the
anti-bunching of fermions. Therefore the amplitude of interference
in the fermionic case should be defined as $A_2=-\overline A$.

Next we want to find the fluctuations of $A$. For this purpose we
need to compute $A_4$ squaring Eq.~(\ref{aq_2}) and substituting
$p(x,t)$ by $n(x,t)$. As we mentioned above we need to normal order
the resulting expression first and then use the asymptotic forms of
the operators $a(x,t)$ and $a^\dagger(x,t)$. As a result one finds
\begin{widetext}
\beq
A_4&=&\int\int\int\int dx_1 dx_2 dx'_1 dx'_2\,
\langle\, : n(x_1,t)n(x_2,t)n(x'_1,t)n(x'_2,t):\,\rangle\,\mathrm e^{iQ(x_1+x_2-x'_1-x'_2)}\nonumber\\
&+&2\int \int\int dx_1 dx_2 dx'_1\, \langle\,: n(x_1,t)n(x_2,t)
n(x'_1,t):\,\rangle\,\mathrm e^{iQ(x_2-x'_1)}+\int\int dx_1
dx_2\,\langle\,:n(x_1,t)n(x_2,t):\,\rangle,
\label{a4_2}
\eeq
where the semicolons imply the normal-ordered form, i.e. that the
creation operators appear on the left of the annihilation operators.
Substituting (\ref{a(t)}) into the integrals above yields
\be
A_4=\langle a_1^{\dagger\,2} a_2^{\dagger\,2} a_2^2\,
a_1^2\rangle+2\,\langle a_1^{\dagger\,2} a_2^\dagger\, a_2\,
a_1^2\rangle+2\,\langle a_1^\dagger a_2^{\dagger\,2} a_2^2\,
a_1\rangle +\langle a_1^{\dagger\,2} a_1^2\rangle+\langle
a_2^{\dagger\,2} a_2^2\rangle+2\langle a_1^\dagger a_2^\dagger a_2
a_1\rangle.
\label{a4_1}
\ee
\end{widetext}

Let us define the relative width of the distribution
$w=\sqrt{A_4-A_2^2}/A_2$. We explicitly look into two different
initial states. First we consider the interference between two
independent coherent states with on average $N$ atoms in each state.
In this case from Eqs.~(\ref{a2}) and (\ref{a4_1}) we find:
\beq
&&A_2=N^2,\; A_4=N^4+4N^3+4N^2\Rightarrow\\
 &&w={2\sqrt{N+1}\over N}.
\eeq
As expected, $w$ vanishes as $N\to\infty$.

Next we consider the interference between the two number states with
$N$ atoms in each of them. Then
\beq
&&A_2=N^2,\; A_4=N^4+2N^3+N^2-2N\Rightarrow\\
&& w={\sqrt{2}\over\sqrt{N}}\sqrt{1+{1\over 2N}-{1\over N^2}}.
\eeq
Asymptotically $w$ also decreases as the inverse square root of the
number of particles at large $N$. However, the coefficient appears
to be smaller by a factor of $\sqrt{2}$ than in the case of the two
coherent states.

We note that Eq.~(\ref{a4_1}) is also valid for fermions. However,
in this case all the terms except the last one identically vanish
and thus $A_4=2A_2^2$ independent of the details of the fermionic
state. This result comes from the fact that in this simple setup
there are at most two interfering particles.

In the case of the interference of extended condensates, which
expand only in the transverse directions (see
Refs.~[\onlinecite{zoran_KT, pnas, gritsev}]), the expressions above
are easily generalized. For example in Eq.~(\ref{a4_1}) one has to
do the following substitutions:
\begin{widetext}
\beq
\langle a_1^{\dagger\,2}a_2^{\dagger\,2}a_2^2\,a_1^2\rangle &\to&
\int\int\int\int dz_1 dz_2 dz_3 dz_4\; \langle a_1^\dagger(z_1)
a_1^\dagger (z_2) a_2^\dagger
(z_3) a_2^\dagger (z_4) a_2(z_1) a_2(z_2) a_1(z_3) a_1(z_4)\rangle,\label{lut1}\\
 \langle a_1^{\dagger\,2}a_2^\dagger a_2\,a_1^2\rangle &\to&
\int\int\int dz_1 dz_2 dz_3\; \langle a_1^\dagger(z_1) a_1^\dagger
(z_2) a_2^\dagger (z_3) a_2(z_2) a_1(z_3) a_1(z_1)\rangle,\\
\langle a_1^{\dagger\,2}\,a_1^2\rangle &\to& \int\int dz_1 dz_2
\langle a_1^\dagger(z_1) a_1^\dagger (z_2) a_1(z_2)
a_1(z_1)\rangle,\\
\langle a_1^\dagger a_2^\dagger a_2\,a_1\rangle &\to& \int\int dz_1
dz_2\; \langle a_1^\dagger(z_1)a_2^\dagger (z_2) a_2(z_1)
a_1(z_2)\rangle.\label{lut4}
\eeq
\end{widetext}
Other substitutions can be obtained from these expressions by simple
permutations. In the equations above $z_j$ denote coordinates along
the condensates and the integration is taken over the area or the
length of the condensates. For the systems with long-range
correlations the terms containing the largest number of bosonic
operators give the leading contribution to $A_4$ in agreement with
the results of Refs.~[\onlinecite{pnas, gritsev}]. Indeed, for
example, for the interference between two 1D condensates at zero
temperature Eqs.~(\ref{a4_1}) and (\ref{lut1}) - (\ref{lut4}) give
\be
A_4=A L^{4-2/K}(1+B (\rho L)^{1/K-1}),
\label{z4_1}
\ee
where $A$ and $B$ are the non-universal constants, $\rho$ is the
mean particle density, $L$ is the system size or the imaging length
(see Refs.~[\onlinecite{pnas, gritsev}] for more details). The
Luttinger parameter $K$ above characterizes the strength of the
interactions in the condensates. For the repulsive bosons with
point-like interactions $K$ interpolates between $1$ in the
impenetrable Tonks-Girardeu regime and $\infty$ in the
noninteracting regime~\cite{cazalilla}. It is clear that shot noise
given by the second term in the brackets of Eq.~(\ref{z4_1}) is
subdominant for large systems as long as $K>1$. We emphasize that
even if shot noise is negligible still $A_4>A_2^2$ as long as $K$ is
finite, i.e. there is no true long range order in the interfering
systems~\cite{pnas}. This inequality implies that $w$ is finite and
the amplitude of the interference fringes still fluctuates because
of nontrivial particle-particle correlation functions in each
system. On the contrary, in systems with short range correlations
the situation changes and one can not simply ignore shot noise. For
example, in one-dimensional condensates at finite temperature $T$
the particle-particle correlation functions decay exponentially at
long distances~\cite{cazalilla}:
\be
\langle a^\dagger (x,t) a(x^\prime,t)\rangle\sim \mathrm
e^{-|x|/\xi_T},
\ee
where $\xi_T\sim 1/T$ is the correlation length. Then instead of
Eq.~(\ref{z4_1}) we find
\be
A_4=\tilde A L^2\xi_T^{2-2/K}\left[1+\tilde B(\rho
\xi_T)^{1/K-1}\right],
\ee
with some other nonuniversal coefficients $\tilde A$ and $\tilde B$.
Note that unlike the zero temperature case the second term, which
corresponds to the shot noise remains finite in the thermodynamic
limit. At low temperatures $\rho\xi_T\gg 1$ its contribution remains
small but the shot noise becomes increasingly important as the
temperature grows and $\xi_T$ decreases. This result can be readily
understood since in systems with finite correlation length only
particles within this length (or the correlation volume in higher
dimensions) coherently contribute into the interference
amplitude~\cite{pnas}. The shorter this length the smaller the
number of such coherent atoms and thus the stronger the effects of
the shot noise.

Next we turn to the interference of multiple independent
condensates. For simplicity we assume that all the condensates are
identical and that they form a one dimensional array with a distance
$d$ separating the nearest neighbors. Then for $M$ condensates the
density operator $n(x,t)$ assumes the following form in the long
time limit:
\be
n(x,t)\approx u^2(x,t)\sum_{j,k=1}^{M-1} a_j^\dagger a_k \mathrm
e^{i(Q_j-Q_k)x},
\label{n_xt}
\ee
where $Q_j=m (x+dj)/\hbar t$ and as before we use the fact that the
Wannier functions corresponding to different condensates are
identical. We note that using Eq.~(\ref{aq_2}) as the measure of the
interference is not very efficient if the number of condensates is
large. In particular, one can show that the relative width of the
distribution of $A_2$ for large $M$ is $w\approx 1+1/N$, where $N$
is the average number of particles per condensate. This width
remains finite even in the limit of large $N$. Physically this
happens because each condensate comes with its own random phase and
thus there is no constructive interference between them. Instead one
can define a different interference measure:
\be
\tilde A\!=\!\sum_q \left[\int\!\int dx dx' p(x,t)p(x'\!,t) \mathrm
e^{iQ_q(x-x')}\!-\!\int\! dx\,p(x,t)\right],
\label{aq_3}
\ee
where $Q_q=q Q$ with $q=1,2,..M-1$. We note that in the case of
multiple condensates there are alternative ways of defining such
measure. For example in Ref.~[\onlinecite{bloch_noise}] F\"olling
{\em et. al.} used the quantity $C(b)=\int dx p(x,t)p(x+b,t).$ The
fluctuations of $C(b)$ can be analyzed by methods similar to those
described in this paper. The evaluation of $\tilde
A_2\equiv\overline{\tilde A}$ is analogous to that of $A_2$ so we
only present the final answer:
\be
\tilde A_2={M(M-1)\over 2}N^2.
\label{a2b}
\ee
Next we can compute $\tilde A_4$, which describes the fluctuations
of $\tilde A_2$. The procedure is again quite similar to one
described above for the case of two condensates. However, the actual
computations and the resulting expression for $\tilde A_4$ are quite
cumbersome, so we present only the two leading terms in the number
of the condensates $M$:
\be
\tilde A_4\approx {M^4 N^4\over
4}+M^3\left(\langle\,:n^2\!:\,\rangle N^2-{5\over 6}N^4+3N^3+{5\over
3}N^2\right)
\ee
This expression yields the following asymptotic form of the relative
width of the distribution of the interference contrast:
\be
\tilde w\approx {2\over N \sqrt{M}}\sqrt{-{N^2\over
3}+\langle\,:n^2:\,\rangle+3N+{5\over 3}}
\label{w1}
\ee
Note that the width $\tilde w$ vanishes with the number of the
condensates even for the small number of particles per condensate.
The coefficient multiplying $1/\sqrt{M}$ in Eq.~(\ref{w1}) weakly
depends on the initial state of each condensate. Thus for the
interference between bosons in number states this coefficient
gradually decreases from $~4.2$ for $N=1$ to $1.6$ for $N\to\infty$.

A very similar calculation can be done for the case of interfering
fermions. The only difference between bosons and fermions is that
all bosonic contributions to $\tilde A_4$ come with a positive sign
and fermionic contributions can bear both negative and positive
signs. While Eq.~(\ref{a2b}) for $\tilde A_2$ remains intact and
holds for fermions as well, the corresponding expression for $\tilde
A_4$ gets modified:
\be
\tilde A_4^{(f)}\approx {M^4 N^4\over 4}+M^3\left(-{3\over
2}N^4+N^3+{1\over 3}N^2\right),
\ee
where the superscript $(f)$ stands for fermions. Correspondingly the
relative width of the distribution becomes:
\be
w^{(f)}\approx {2\over N \sqrt{M}}\sqrt{-N^2+N+{1\over 3}}.
\label{w2}
\ee

Note that both Eqs.~(\ref{w1}) and (\ref{w2}) have similar scaling
with the number of the interfering systems. This observation leads
us to an interesting conclusion that a high interference contrast is
possible for the case of multiple incoherent fermionic sources. This
is opposite to the situation of two interfering systems, where the
fringe contrast always remains low.

We would like to mention that in real experiments dependence of the
interference contrast on $M$ can saturate at large $M$. The reason
is that the derivation of Eqs.~(\ref{w1}) and (\ref{w2}) relied on
taking into account all the Fourier components appearing in
Eq.~(\ref{aq_3}). However, at large wave vectors $q_Q$ this can
become problematic because of finite resolution of the apparatus,
finite width of Wannier functions, not entirely free expansion of
the atoms etc. Thus realistically the sum over $q$ in
Eq.~(\ref{aq_3}) is always bounded by some value $q_{max}$. Then
obviously increasing the number of interfering systems beyond
$q_{max}$ will not improve the interference contrast. We also point
that Eq.~(\ref{n_xt}) is valid only in the long time limit provided
that $\hbar^2 t/m\gg (Md)^2$, or equivalently $w\gg (Md)^2/w_0$,
where $w$ is the width of the condensate after the expansion and
$w_0<d$ is the initial width of each condensate. This condition is
hard to achieve if the number of the interfering systems is large.
Alternatively one can view the relation $M^\star=\sqrt{w w_0}/d$ as
determining the maximum number of the interfering systems beyond
which the contrast saturates.

In conclusion, we analyzed fluctuations of the amplitude of
interference of independent atomic systems. We showed that for the
interference of two bosonic systems the fluctuations of the fringe
amplitude are inversely proportional to the square root of the
number of particles in each system. The coefficient of
proportionality is non-universal; it depends on the details of the
wave functions (or more generally density matrices) of each system.
In particular, we found that the fluctuations are larger for the
case of two interfering condensates with broken phases than for the
case of two Fock states. We generalized our analysis to the
interference of multiple systems and showed that fluctuations of the
fringe contrast vanish as the number of interfering systems (bosonic
or fermionic) becomes large.

The author would like to acknowledge E.~Altman, J.~Dalibard,
E.~Demler, Z.~Hadzibabic, A.~Imambekov, and V.~Gritsev for useful
discussions. This work was supported by AFOSR YIP.

\end{document}